\documentclass[11pt,a4paper]{article}
\usepackage{jheppub}

\title{\boldmath Matter Effect of Light Sterile Neutrino: An Exact Analytical Approach}

\author[a]{Wei Li}
\author[b]{Jiajie Ling}
\author[a]{Fanrong Xu\footnote{Corresponding author.}}
\author[b]{Baobiao Yue}

\affiliation[a]{Department of Physics and Siyuan Laboratory, Jinan University,\\
$\;$Guangzhou 510632, P.R. China}
\affiliation[b]{School of Physics, Sun Yat-Sen University,\\
$\;$Guangzhou 510275, P.R. China}

\emailAdd{weili@stu2014.jnu.edu.cn}
 \emailAdd{ lingjj5@mail.sysu.edu.cn}
 \emailAdd{ fanrongxu@jnu.edu.cn}
  \emailAdd{yuebb@mail2.sysu.edu.cn.}

\abstract{
The light sterile neutrino, if it exists, will give  additional contribution to matter effect when active neutrinos propagate through
terrestrial matter.
In the simplest 3+1 scheme, three more rotation angles and two more CP-violating phases in lepton mixing matrix make the 
interaction complicated formally.
 In this work, the  exact analytical expressions for active neutrino oscillation probabilities 
 in terrestrial matter, including sterile neutrino contribution,  are derived. 
 It is pointed out that this set of formulas contain information
 both in matter and in vacuum, and can be easily tuned by choosing related parameters.
 Based on the generic exact formulas, we present oscillation probabilities of typic medium and long baseline experiments.
 Taking NO$\nu$A experiment as an example, we show that in particular parameter space sterile neutrino gives
 important contribution to terrestrial matter effect, and Dirac phases play a vital role.
  
 }


\begin{document}
\maketitle
\flushbottom


\section{Introduction}
\label{sec:intro}	

In Standard Model (SM) as a component of SU(2) left-handed doublet, neutrino is electric neutral, massless and
only takes part in weak interaction.
Now it has been well established that at least two active neutrinos are massive with tiny masses. 
The origin of neutrino mass is still an open question. 
Including seesaw mechanism \cite{seesaw} and radiative correction mechanism (for example,\cite{Kohda:2012sr}, \cite{Chang:2016zll} and for a recent review see \cite{Cai:2017jrq})
many efforts have been contributed to this . 
General speaking, new particles out of SM particle spectrum will appear associated with neutrino mass models.
As a hypothetic particle, though does not participate weak interaction, sterile neutrino 
is required in some neutrino mass models beyond SM.
For example, in Type I seesaw mechanism the heavy right-handed neutrino singlet contributing
the tiny mass of left-handed neutrino is  absent from SU(2) interaction
and hence  is  the sterile neutrino. However, 
the mass of sterile neutrino, varied from eV to TeV,  has not been determined yet.

Recently the search of sterile neutrino in experiment is active. 
For heavy sterile neutrino, the run of LHC provides a particular opportunity.
The IceCube Neutrino Observatory, which locates in Antarctic, gives an unique vision.
And recently, an event of high energy neutrino which is absolutely beyond the structure of SM, has been
observed by IceCube \cite{IceCube:2018}.  
Meanwhile neutrino oscillation experiments are usually considered
as a useful platform to extract information of light sterile neutrino.
Indeed it has  been implicated by oscillation experiments the existence of sterile neutrino.
In 2001 the LSND experiment searched $\bar{\nu}_\mu\to\bar{\nu}_e$ oscillations,
suggesting that neutrino oscillations occur in the $0.2<\Delta m^2<10\, {\rm{eV}}^2$  range \cite{Aguilar:2001ty}.
Later the MiniBooNE experiment indicated a two-neutrino oscillation, $\bar{\nu}_\mu\to\bar{\nu}_e$,
occurred in the $0.01<\Delta m^2<1.0\,{\rm{eV}}^2$ range \cite{Aguilar-Arevalo:2013pmq}.
Recently combining $\bar{\nu}_e$ disappear mode from  Daya Bay  Collaboration and
$\nu_\mu$ neutrino oscillation from MINOS, the two Collaborations give a joint analysis,
incorporating  the early Bugey-3 data, and claim that at $95\%$ C.L.
$\Delta m_{41}^2<0.8 {\rm{eV}}^2$ can be excluded \cite{Adamson:2016jku}.
The IceCube neutrino telescope, measuring the atmospheric muon neutrino spectrum,
extend the exclusion limits to $\sin^22\theta_{24}\leq 0.02$ at $\Delta m^2\sim 0.3 {\rm{eV}}^2$
at $90\%$ confidence level in 2016 \cite{TheIceCube:2016oqi}.
From a recent effort of NEOS collaboration, the mixing parameter limit is obtained as $\sin^22\theta_{14}\leq 0.1$
for $0.2<\Delta m_{41}^2<2.3 {\rm{eV}}^2$  \cite{Ko:2016owz}.
Taking into account recent progress, 
a global fit of short-baseline neutrino oscillation has been 
updated \cite{Gariazzo:2017fdh}, giving $\Delta m_{41}^2\approx 1.7  {\rm{eV}}^2$ (best-fit), $1.3  {\rm{eV}}^2$ (at 2$\sigma$),
$2.4  {\rm{eV}}^2$ (at $3\sigma$) and $0.00047\leq \sin^2 2\theta_{e\mu}\leq 0.0020$ at $3\sigma$.
Sterile neutrino, probably as a port to new physics, is far more than clear today.

When neutrino propagates in matter,  the matter effect should be taken into account and
the significance is different. It is known that usually in short and medium baseline 
experiments, matter effect does not give a dominated contribution while in long baseline experiment,
oscillation could be largely affected by matter effect. However, in the precise experiment like JUNO, 
even though the baseline is not so long, the matter effect probably reduce the sensitivity of mass ordering 
measurement, thus a careful study of terrestrial matter effect on medium baseline experiment is
performed \cite{Li:2016txk}.
A similar analysis should be considered if sterile neutrino exists regardless the length of experiments' baseline.

Some efforts have been contributed.
Klop and Palazzo studied sterile neutrino induced  CP violation with T2K data  \cite{Klop:2014ima},
where they developed an approximated method and helps to simplify the calculation.
Choubey {\it et. al.} extended the discussion to DUNE, T2HK and T2HHK experiments \cite{Choubey:2017cba}, \cite{Choubey:2017ppj}.
In \cite{Ghosh:2017atj}, Ghosh {\it et. al.} studied mass hierarchy sensitivity in presence of sterile neutrino in NO$\nu$A.
A general discussion on light sterile neutrino effect in long baseline experiments has been performed by Dutta {\it et. al.}\cite{Dutta:2016glq}.
Later a joint short- and long-baseline constraints on light sterile neutrino have been done by Capozzi {\it el. al.} \cite{Capozzi:2016vac}.
Recently, more works related to T2HK, DUNE and NO$\nu$A have been contributed in  \cite{Palazzo:2018},\cite{Chatla:2018sos} and \cite{Gupta:2018qsv}, 
and an updated global analysis  has been done by Dentler {\it el. al.} \cite{Dentler:2018sju}.
In addition to the approximated method proposed by Klop and Palazzo, some other alternative  approach has also been proposed
\cite{LLXY2018}.

Meanwhile we should keep in mind that in above works, the analysis is based on either approximate analytical method or numerical approach.
For the particular modes, such approach is convenient. On the other hand,
since the sterile neutrino mass (even light sterile neutrino mass) is unknown,
the approximation adopted  above has a risk to lose some information though  calculation speeds up.
Thus a complete exact analytical solution, formally complicated, is valuable.  
Such efforts were made previously in \cite{Zhang:2006yq} and \cite{Kamo:2002sj}, in two different 
approaches.
In this work, we will develop the method and improve the result in \cite{Zhang:2006yq}, and then provide  a complete exact analytical solution.

This paper is organized as follows. In section \ref{Oscillation} we will give a brief introduction of neutrino oscillation with
the consideration of matter effect. 
In section \ref{sec:LSNwithME} we will 
derive the mass-square differences within matter effect, show related
rotation matrix elements and propagation probabilities explicitly.
The applications of such analytical solution will be presented in section \ref{sec:application}.
In section \ref{sec:conclusion}, we will summarize and give a conclusion. More details
involved in section \ref{sec:LSNwithME} are shown in appendix.

\section{Theorectial Framework}\label{Oscillation}

The picture of neutrino oscillation is well understood currently. 
The identity of neutrinos in flavor space and mass space is not identical, or they have a mixing.
Due to such a mixing, described by rotation matrix $U$, the identity of neutrinos can be changed during its journey from source to destination, called
neutrino oscillation. The oscillation probability, 
that is the probability for capturing neutrino as $\nu_\beta$ from the initial beam $\nu_\alpha$,  is
\begin{equation}
P(\nu_\alpha\to\nu_\beta)=\sum_i|U_{\alpha i}|^2 |U_{\beta i}|^2 + 2 \sum_{i<j}
\left[{\rm{Re}}(U_{\alpha i} U_{\beta j} U_{\alpha j}^* U_{\beta i}^*)\cos\Delta_{ij}
-{\rm{Im}}(U_{\alpha i} U_{\beta j} U_{\alpha j}^* U_{\beta i}^*) \sin\Delta_{ij}
\right],\label{eq:P}
\end{equation}
where $\Delta_{ij}\equiv \Delta m_{ij}^2 L/(2E)$ with $\Delta m_{ij}^2=m_i^2-m_j^2$,  while $L$  is 
propagating distance, and
 $E$ is the energy carried by neutrinos. Both appear mode and disappear mode are contained in Eq. (\ref{eq:P}).
 
 The oscillation probability is determined by universal parameters $U_{\alpha i}$,
 $m_i$ as well as experiment dependent parameters $E$ and $L$. In Standard Model (SM) there are only three 
 flavors of active neutrinos, thus the mixing matrix, named PMNS matrix, is parameterized by three rotation angles 
 and one CP-violating phase. Within this theoretical framework the three angles ($\theta_{12}, \theta_{23}, \theta_{13}$)
 are measured by solar neutrino, atmospheric neutrino and reactor neutrino experiments, respectively.
 The remaining undetermined parameter is CP phase $\delta$, as well as the sign of $\Delta m_{13}^2$,
 could be reachable in the following ten years. 
 
 On the other hand, the possibility to have one more light sterile neutrino 
 still exists. 
The sterile neutrino (denoted as $\nu_s$), unlike the active neutrinos 
(denoted as $\nu_e, \nu_\mu, \nu_\tau$ in flavor state),
 are known for its absence from SM weak interactions. 
 However, its  effect appears indirectly by 
mixing with active neutrinos.
Such a mixing is described by lepton mixing matrix $U$, 
 given
\begin{equation}
\left(
\begin{array}{c}
\nu_e\\ \nu_\mu\\ \nu_\tau \\ \nu_s
\end{array}\right)
=\left(\begin{array}{cccc}
U_{e1} & U_{e2} & U_{e3} & U_{e4}\\
U_{\mu1} & U_{\mu 2} & U_{\mu 3} & U_{ \mu 4}\\
U_{\tau 1} & U_{\tau 2} & U_{\tau 3} & U_{\tau 4}\\
U_{s1} & U_{s2} & U_{s3} & U_{s4}
\end{array}
\right)
\left(
\begin{array}{c}
\nu_1\\ \nu_2\\ \nu_3 \\ \nu_4
\end{array}\right),
\end{equation}
which characterizes the rotation between mass eigenstate and flavor eigenstate in vacuum.
With more degrees of freedom in $U$, the oscillation probability will contain richer information.
There are 6 angles and 3 phases to parameterize mixing matrix $U$. 
Similar to the standard parameterization of PMNS matrix, putting the two extra phases in 1-4, 3-4 plane,
we can write down the four dimensional mixing matrix as
\begin{equation}
U=R(\theta_{34},\delta_{34})R(\theta_{24})R(\theta_{14},\delta_{14})R(\theta_{23})R(\theta_{13},\delta_{13})R(\theta_{12}),
\end{equation}
in which $R\left(\theta_{ij}(,\delta_{ij})\right)$ represents Euler rotation in i-j plane without (with) a CP phase.
More details for four dimensional $U$ are shown explicitly in appendix \ref{sec:app-mix}. 

When passing through matter, active neutrinos interact with matter by weak interaction. More exactly $\nu_e$ 
interacts via both charged current and neutral current while $\nu_\mu, \nu_\tau$ only receive neutral current interaction
by exchanging $Z$ bosons.  Though sterile neutrino itself does not take part in weak interaction, by removing the global neutral current which
will not affect oscillation probability,
$\nu_s$ has an induced nonzero term in effective Hamiltonian while the corresponding ones for $\nu_{\mu, \tau}$ vanish, given
\begin{equation}
\tilde{H}_{\rm{eff}}
=\frac{1}{2E} \left[ {U}\left(\begin{array}{cccc}
{m}_1^2 & 0 & 0 & 0 \\
0 & {m}_2^2  & 0 & 0 \\
 0 & 0& {m}_3^2  & 0 \\
 0 & 0 & 0& {m}_4^2
\end{array}\right) {U}^\dagger +
\left(\begin{array}{cccc}
A & 0 & 0 & 0 \\
0 & 0  & 0 & 0 \\
 0 & 0& 0  & 0 \\
 0 & 0 & 0& A'
\end{array}\right) \right], \label{effectiveH}
\end{equation}
where $U$ is the lepton mixing matrix in vacuum and $A=2\sqrt{2} G_F N_e E$,  $A'=-\sqrt{2} G_F N_n E$
with densities of electron (neutron) $N_{e}(N_{n})$.  Without loss of generality, the Hamiltonian can always 
be written in a more compact form
\begin{equation}
\tilde{H}_{\rm{eff}}=\frac{1}{2E} \left[ \tilde{U}\left(\begin{array}{cccc}
\tilde{m}_1^2 & 0 & 0 & 0 \\
0 & \tilde{m}_2^2  & 0 & 0 \\
 0 & 0& \tilde{m}_3^2  & 0 \\
 0 & 0 & 0& \tilde{m}_4^2
\end{array}\right) \tilde{U}^\dagger \right]\label{effectiveH2}
\end{equation}
where the effective mass $\tilde{m}_i$ and the new defined effective lepton mixing matrix $\tilde{U}$ have incorporated  information 
of matter effect. And hence the oscillation probability including matter effect has the same structure
of the one in vacuum, that is
\begin{equation}
\tilde{P}(\nu_\alpha\to\nu_\beta)=\sum_i|\tilde U_{\alpha i}|^2 |\tilde U_{\beta i}|^2 + 2 \sum_{i<j}
\left[{\rm{Re}}(\tilde U_{\alpha i} \tilde U_{\beta j} \tilde U_{\alpha j}^* \tilde U_{\beta i}^*)\cos\tilde \Delta_{ij}
-{\rm{Im}}(\tilde U_{\alpha i} \tilde U_{\beta j} \tilde U_{\alpha j}^* \tilde U_{\beta i}^*) \sin\tilde\Delta_{ij}
\right],\label{eq:P-mass}
\end{equation}
with $\tilde \Delta_{ij}\equiv \Delta \tilde m_{ij}^2 L/(2E)$, $\Delta\tilde m_{ij}^2=\tilde m_i^2-\tilde m_j^2$.
Hereafter we will adopt ${P}$ to stand for $\tilde{P}$ for its clear meaning. 
Obviously if one works out explicitly $\tilde{U}_{\alpha i}$ and $\tilde{m}_i^2$, the probability will be well presented.
However, such a calculation would be challenging.
To avoid  the difficulty, 
by working out the effective mass differences and some necessary combinations 
of entries of $\tilde{U}$, we can  obtain  complete exact expressions for $P$ as well.
In the following section, we will derive the necessary parameters.

\section{Effective Parameters}\label{sec:LSNwithME} 
Mass difference $\Delta m_{ij}^2$ and lepton mixing matrix $U_{\alpha i}$  in vacuum are universal. 
The corresponding ones in matter will be corrected by matter parameters.
We will give $\Delta \tilde m_{ij}^2$ firstly, based on which $\tilde U_{\alpha i}$ is also obtained.

\subsection{Effective mass-square difference}

In this part, we aim at solve $\Delta \tilde{m}^2_{ij}$.
It is known that a constant  can be removed from diagonal entries simultaneously,
as it contributes to a global phase which does not affect oscillation probability.
Then by subtracting a global $m_1^2$ in Eq. (\ref{effectiveH}), we have
\begin{equation}
\left[ \tilde{U}\left(\begin{array}{cccc}
\hat{\Delta}{m}_{11}^2 & 0 & 0 & 0 \\
0 & \hat{\Delta}{m}_{21}^2  & 0 & 0 \\
 0 & 0& \hat\Delta{m}_{31}^2  & 0 \\
 0 & 0 & 0& \hat\Delta{m}_{41}^2
\end{array}\right) \tilde{U}^\dagger \right]
=\left[ {U}\left(\begin{array}{cccc}
0 & 0 & 0 & 0 \\
0 & \Delta{m}_{21}^2  & 0 & 0 \\
 0 & 0& \Delta{m}_{31}^2  & 0 \\
 0 & 0 & 0& \Delta{m}_{41}^2
\end{array}\right) {U}^\dagger +
\left(\begin{array}{cccc}
A & 0 & 0 & 0 \\
0 & 0  & 0 & 0 \\
 0 & 0& 0  & 0 \\
 0 & 0 & 0& A'
\end{array}\right) \right]
\label{eq:diagH}
\end{equation} 
in which we have defined $\hat\Delta m_{ij}^2=\tilde{m}_i^2-m_j^2$. 
The induced mass  difference can also be written in the form of $\hat{\Delta}m_{ij}^2=\hat{\Delta}m_{i1}^2-\Delta m_{j1}^2$.
With the help of $\hat\Delta m_{ij}^2$, the effective mass difference
$\Delta \tilde{m}^2_{ij}$ which we seek for is constructed as
\begin{equation}
{\Delta}\tilde m_{ij}^2= \hat{\Delta}m_{i1}^2-\hat\Delta m_{j1}^2.
\label{eq:effMD}
\end{equation}
The effective mass difference between two arbitrary effective masses can
be resort to those differences from $\tilde{m}_1^2$.
Thus the aim now is simplified to find out $\hat{\Delta} m_{i1}^2$, that is  the
diagonalization of the right-handed side of Eq.(\ref{eq:diagH}).

In principle the key point of the diagonalization is to solve a quartic equation, which is fortunately solvable. 
The particular involved quartic equation as well as its solutions are shown in appendix \ref{subsec:eigenequation}.
With necessary new-defined parameters, 
we can obtain
the exact analytical expressions for $\hat{\Delta}m_{j1}^2 \, (j=1,2,3,4)$, 
\begin{align}
& \hat{\Delta} m_{11}^2=-\frac{b}{4}-S-\frac12\sqrt{-4S^2-2p+\frac{q}{S}}\label{eq:hatDelta}\\
&\hat{\Delta} m_{21}^2=-\frac{b}{4}-S+\frac12\sqrt{-4S^2-2p+\frac{q}{S}}\nonumber\\
& \hat{\Delta} m_{31}^2=-\frac{b}{4}+S-\frac12\sqrt{-4S^2-2p-\frac{q}{S}}\nonumber\\
& \hat{\Delta} m_{41}^2=-\frac{b}{4}+S+\frac12\sqrt{-4S^2-2p-\frac{q}{S}},\nonumber
\end{align}
which depend on ${\Delta}m_{i1}^2 \, (i=1,2,3,4)$.
Hence by the usage of Eq. (\ref{eq:effMD}) the effective mass difference, including matter effect correction, now 
can be expressed explicitly
\begin{align}\label{massdiff}
&{\Delta}\tilde m^2_{21}=\sqrt{-4S^2-2p+\frac{q}{S}},\\
&{\Delta} \tilde m^2_{31}= 2S +\frac12\left(\sqrt{-4S^2-2p +\frac{q}{S}}
-\sqrt{-4S^2-2p -\frac{q}{S}}\right)\,\nonumber\\
&{\Delta}\tilde m^2_{41}=2S+\frac12\left(\sqrt{-4S^2-2p+\frac{q}{S}}
+\sqrt{-4S^2-2p-\frac{q}{S}}\right).\nonumber
\end{align}
in which $b, p, q$ and $S$ are intermediate parameters defined in appendix \ref{subsec:eigenequation}.
Note here we have assumed normal mass hierarchy $(m_1^2<m_2^2<m_3^2<m_4^2)$. 
Without loss of generality, other situations of mass ordering can be derived similarly.

\subsection{Effective lepton mixing matrix}

In addition to effective mass difference, the oscillation probabilities of neutrino propagation rely on lepton mixing matrix as well.
Without relating to each entry of the matrix, only some particular combinations are concerned.
We have shown how to solve these quantities in Appendix \ref{app:mixing}, in which
the general expressions  have been given in Eq. (\ref{U234}, \ref{U1},  \ref{eq:F}, \ref{eq:C}).
In this section, we restrict our interests typically in reactor neutrino  and accelerator neutrino experiments.
The relevant entries are listed below explicitly.

\begin{itemize}

\item The reactor neutrino experiments: $\bar{\nu}_e\to \bar{\nu}_e$

For the disappear mode of anti-electron  neutrino, the only concerned entry  $\tilde{U}_{ei}$ is
\begin{equation}
|\tilde{U}_{e i}|^2
=
\frac{1}{\prod\limits_{k\neq i} {\Delta}\tilde m_{ik}^2} \left(X_e +C_e\right), 
\end{equation}
in which the auxiliary quantities are
\begin{align}
&X_e=\sum_j F_e^{ij}|U_{ej}|^2,\quad
F_e^{ij}=\prod_{k\neq i}\left(A+\hat{\Delta} m_{jk}^2\right)\nonumber\\
& C_e = -A \sum_{i<j} (\Delta m_{ij}^2)^2|U_{ei}|^2 |U_{ej}|^2 - A' \sum_{i<j}
(\Delta m_{ij}^2)^2{\rm {Re}}( U_{ei}U_{ej}^* U_{si}^* U_{sj}).
\end{align}
For the first glance, $|\tilde{U}_{e i}|^2$ relies on $\Delta\tilde{m}^2_{ij}, \Delta m^2_{ij}$ and $U_{\alpha i}$.
Note $\Delta\tilde{m}^2_{ij}=\hat{\Delta}m_{i1}^2-\hat{\Delta}m_{j1}^2$ and $\hat{\Delta}m_{i1}^2$
is the solution for quartic equation which further relies on $\Delta m^2_{i1}$ and $U_{\alpha i}$ as 
well as matter effect parameters $A, A'$. 
So the free parameters for  matter effect correction to $|\tilde{U}_{e i}|^2$ 
are $\Delta m^2_{ij}, U_{\alpha i}, A$ and $A'$.

\item  The accelerator neutrino experiment: disappear mode $\nu_\mu\to \nu_\mu$

Both disappear mode and appear mode will be used in accelerator neutrino experiments.
For the  disappear mode,
the required $|\tilde{U}_{\mu i}|^2$ is given as
\begin{equation}
|\tilde{U}_{\mu i}|^2
=
\frac{1}{\prod\limits_{k\neq i} {\Delta}\tilde m_{ik}^2} \left(X_\mu +C_\mu\right),
\end{equation}
with the associated functions
\begin{align}
&X_\mu=\sum_j F_\mu^{ij}|U_{\mu j}|^2,\quad
F_\mu^{ij}=\prod_{k\neq i} \hat{\Delta} m_{jk}^2 \nonumber\\
& C_\mu = -A \sum_{i<j} (\Delta m_{ij}^2)^2 {\rm{Re}}(U_{\mu i} U_{\mu j}^* U_{ei}^* U_{ej})
-A'\sum_{i<j}(\Delta m_{ij}^2)^2
{\rm{Re}}
(U_{si} U_{sj}^*
U_{\mu i}^* U_{\mu j}).
\end{align}
Except a difference in $F_\alpha^{ij}$, all other terms are same as $|\tilde{U}_{ei}|^2$ up to a 
corresponding change $e\to \mu$. One may find a consistent result from
\cite{Zhang:2006yq}.

\item  The accelerator neutrino experiment: appear mode $\nu_\mu\to \nu_e$

In this case, a distinct difference from disappear mode is that the
product of two entries, $\tilde{U}_{e i} \tilde{U}^*_{\mu i}$,  are required. 
One can immediately have the following relation according to
the general expression in Appendix \ref{app:mixing},
\begin{equation}
\tilde{U}_{e i} \tilde{U}^*_{\mu i}=\frac{1}{\prod\limits_{k\neq i} {\Delta}\tilde m_{ik}^2}
\left[ \sum_j F^{ij}_{e \mu} U_{ej} U_{\mu j}^* + C_{e\mu}\right]
\end{equation}
associated with
%
\begin{align}
& F_{e\mu}^{ij}=\Bigg[A^2\Delta m_{j1}^2+A\Delta m_{j1}^2\Big(\Delta m_{j1}^2-\sum_{k\neq i}\hat{\Delta}m_{k1}^2\Big)+(\Delta m_{j1}^2)^3
-\sum_{k\neq i}(\Delta m_{j1}^2)^2\hat{\Delta}m_{k1}^2
\notag\\
&\hspace{1.5cm}+\sum_{k,l;k\neq l\neq i}\Delta m_{j1}^2\hat{\Delta}m_{k1}^2\hat{\Delta}m_{l1}^2\Bigg]\notag\\
& C_{e\mu}=A'\sum_{k,l}\Delta m_{k1}^2\Delta m_{l1}^2U_{ek}U_{\mu l}^*U_{sk}U_{sl}^*+A\sum_{k,l}\Delta m_{k1}^2\Delta m_{l1}^2
|U_{ek}|^2U_{el}U_{\mu l}^*
\end{align}
Note the corresponding result provided in \cite{Zhang:2006yq} is not consistent with ours, while our calculation can be confirmed by 
 numerical evaluation.
  
\end{itemize}

\subsection{Exact oscillation probability}

Armed with effective mass difference and effective mixing matrix entries, the 
oscillation probabilities are spontaneously presented as,
\begin{equation}
P(\bar{\nu}_e\to\bar{\nu}_e)=1- 4 \sum_{i<j} \left(|\tilde{U}_{ei}|^2|\tilde{U}_{ej}|^2
\sin^2\frac{\tilde{\Delta}_{ij}}{2}\right),
\end{equation}

\begin{equation}
P(\nu_\mu\to \nu_\mu)=1-4\sum_{i<j} \left(
|\tilde{U}_{\mu i}|^2|\tilde{U}_{\mu j}|^2 \sin^2\frac{\tilde{\Delta}_{ij}}{2}\right),
\end{equation}

\begin{equation}
P(\nu_\mu\to \nu_e) 
=
\sum_i|\tilde{U}_{\mu i}|^2 |\tilde{U}_{e i}|^2+
2\sum_{i<j} \left[
{\rm{Re}} (\tilde{U}_{e i} \tilde{U}_{\mu j} \tilde{U}^*_{e j}
\tilde{U}^*_{\mu i}) \cos\tilde{\Delta}_{ij}
-{\rm{Im}} (\tilde{U}_{e i}\tilde{U}_{\mu j}
\tilde{U}^*_{e j} \tilde{U}^*_{\mu i})\sin\tilde{\Delta}_{ij}
\right],
\end{equation}
where  $\tilde{\Delta}_{ij}\equiv\Delta \tilde{m}^2_{ij}L/2E$ and $L$ is
the baseline of a particular neutrino experiment.
The input parameters are (${\Delta} m^2_{i1}$, $U_{\alpha i}$, $A$, $A'$),
where the description of $U_{\alpha i}$ further relies on their parametrization, one
of them can be found in Appendix \ref{sec:app-mix}.

Thorough out the whole derivation, no additional assumptions
are adopted  except the unitary condition of $U$ and $\tilde{U}$.
So the exact analytical expressions are applicable for both short baseline and long baseline
experiments.
Meanwhile we would like to point out that the formulas derived here are the most generic ones 
in 3+1 scheme, since all possible situations, including SM case, are all contained in. In particular, 
we may get the following extreme cases by tuning parameters in our formulas, 
\begin{itemize}
\item[i)]  active neutrino propagating in matter with sterile neutrino effect:  $U_{\alpha i}\neq 0, A\neq 0, A'\neq 0$.
\item[ii)] active neutrino propagating in vacuum with sterile neutrino effect: $U_{\alpha i}\neq 0, A=0, A'=0$.
\item[iii)] active neutrino propagating in matter without sterile neutrino effect: $U_{\alpha 4}= 0, U_{si}=0(, A'= 0),  A\neq 0$,
in which whether $A'$ vanishes doesn't give an effect.
\item[iv)] active neutrino propagating in vacuum without sterile neutrino effect: $U_{\alpha 4}= 0, U_{si}=0(, A'= 0),  A=0$.
\end{itemize}
By setting $\theta_{14}=\theta_{24}=\theta_{34}=0$, 
to close parameters $U_{\alpha 4}$ and $U_{si}$
can be easily fulfilled.

\section{Applications and Discussion}
\label{sec:application}

The exact analytical solution keeps the original information of sterile neutrino without any approximation.
Since sterile neutrino mass is still unknown, approximated formulas, though can speed up evaluation, 
still have a risk to lose some information.
In this section, based on the exact solutions, we give a numerical analysis for typical neutrino experiments.

For each experiment, two types of input parameters are relevant.
One type is the universal parameters, including mixing matrix and mass difference, while the other non-universal one
depends on experiment location, neutrino source and matter effect parameters. For illustration, we take input parameters as follows.
There are 6 rotation angles and 3 Dirac phases in mixing matrix, while the oscillation irrelevant Majorana phases can be ignored here.
 We take $\sin^2\theta_{13}=0.0218, \sin^2\theta_{12}=0.304, \sin^2\theta_{23}=0.437$
 from the SM global fitting\cite{Gonzalez-Garcia:2014bfa}, 
 the other 3 angles we choose
$\sin^2\theta_{14}=0.019, \sin^2\theta_{24}=0.015, \sin^2\theta_{34}=0$ \cite{Gariazzo:2017fdh}. 
Throughout the simulation, we fix one of the three Dirac phases as $\delta_{34}=0$, and let
the other two as free parameters for the purpose of illustration.
As for the mass-square difference, two of the three are consistent with SM global fitting,
$\Delta m_{21}^2=7.5\times 10^{-5} {\rm{eV}}^2, \Delta m_{31}^2=2.457\times 10^{-3} {\rm{eV}}^2$, the remaining one is fixed as
$\Delta m_{41}^2=0.1 {\rm{eV}}^2$. 
To describing matter effect, we adopt the relevant parameters from realistic oscillation experiment \cite{Mocioiu:2000st}, 
which set matter density as $\rho\approx2.6g/cm^3$ and eletron fraction $Y_e\approx0.5$.

\subsection{Medium baseline experiment}

\begin{figure}[h]
\begin{center}
\begin{tabular}{cc}
\includegraphics*[width=0.45\textwidth]{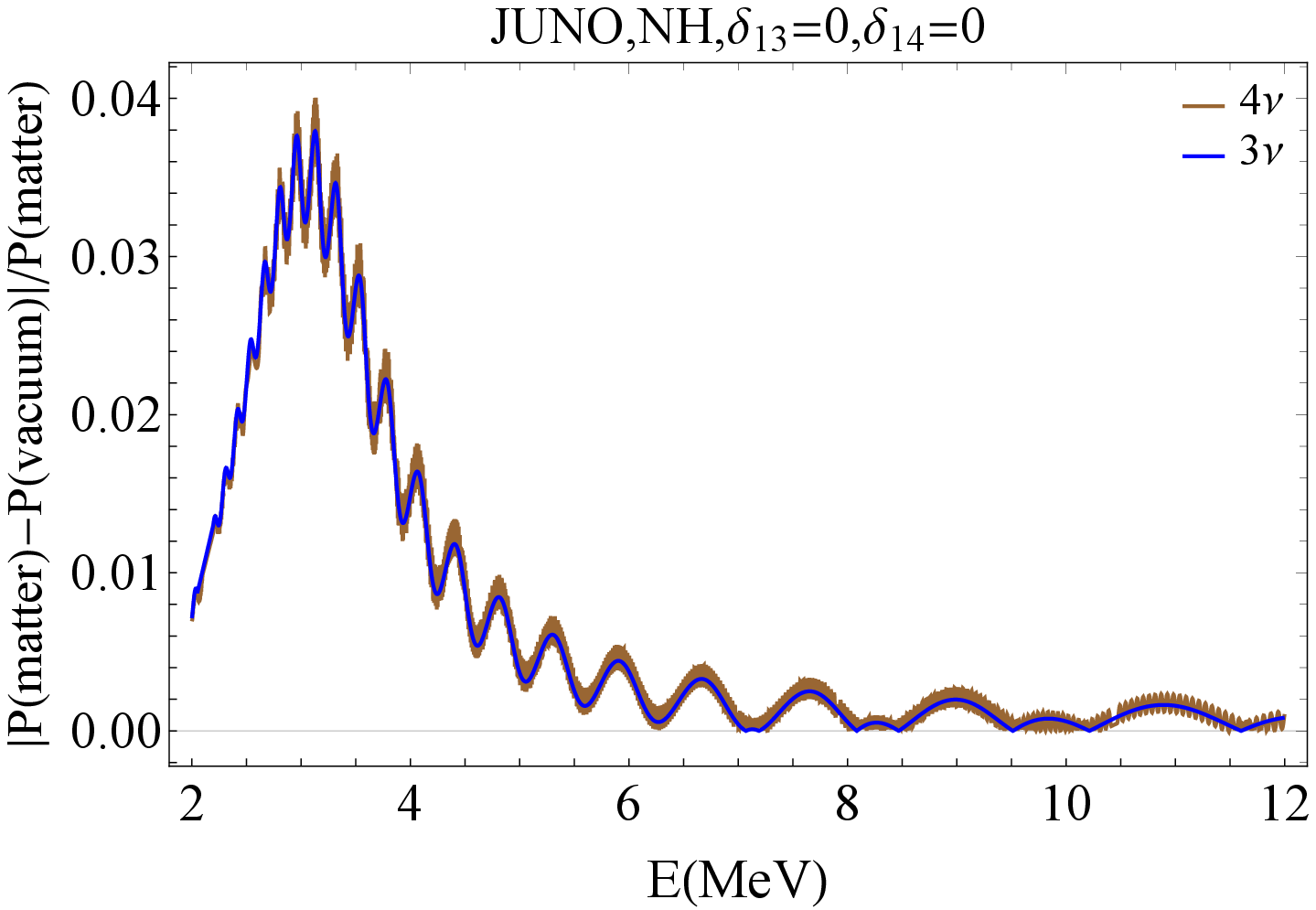}
&
\includegraphics*[width=0.45\textwidth]{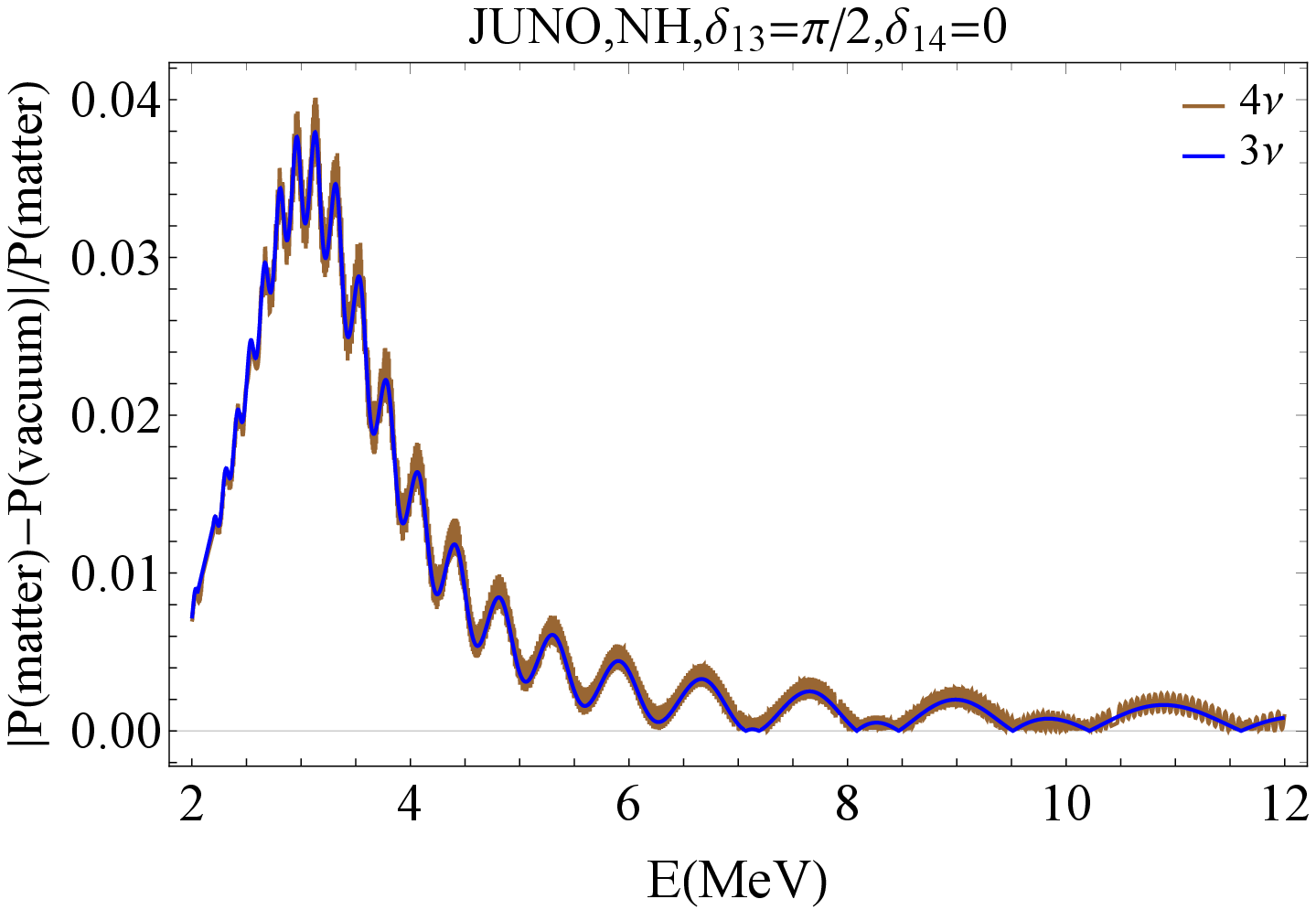}
\\
(a) & (b)\\
\\
\includegraphics*[width=0.45\textwidth]{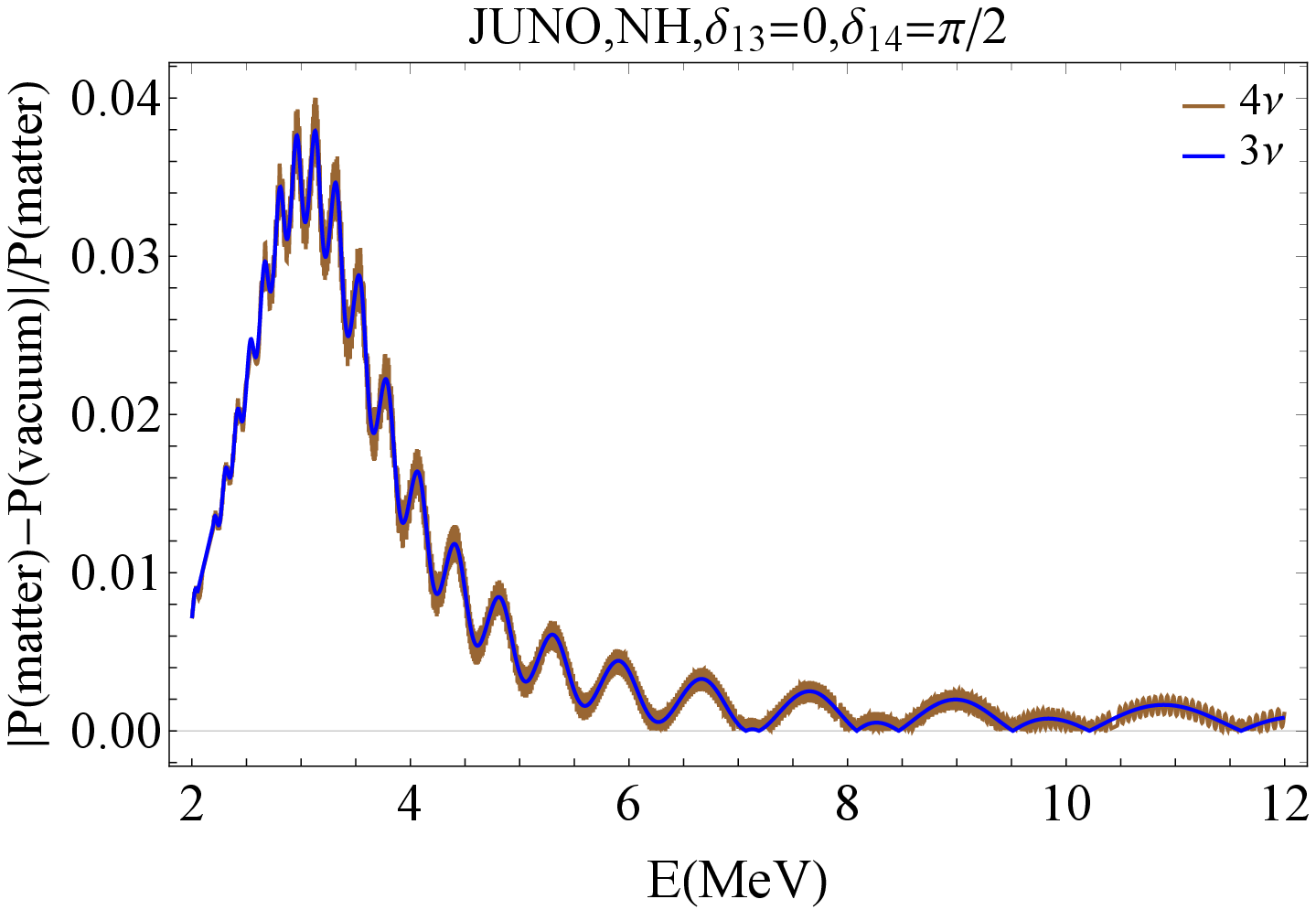}
&
\includegraphics*[width=0.45\textwidth]{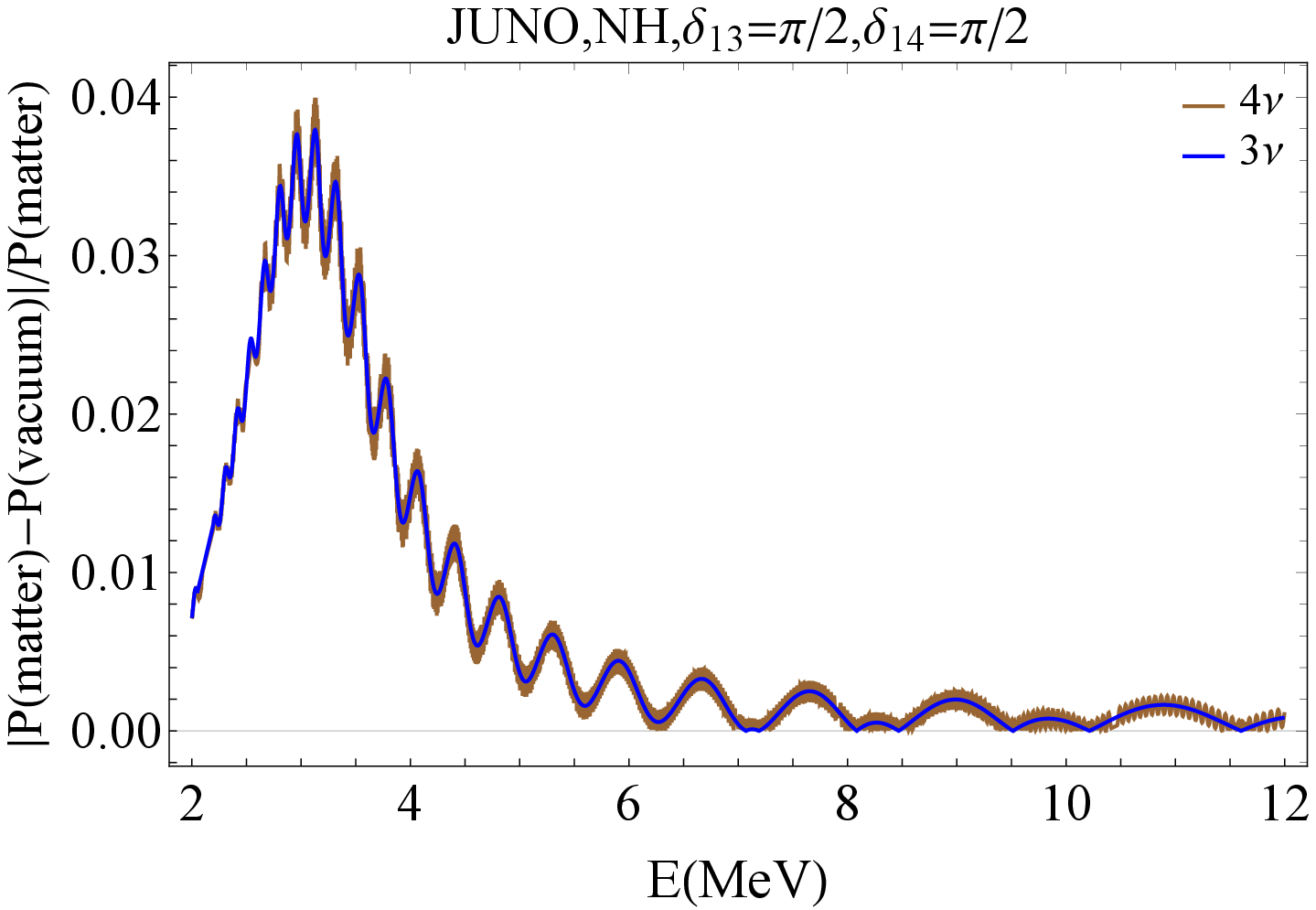}\\
(c) & (d) \\
\end{tabular}
\newline
\vspace{-0.6cm} \caption{Relative difference for the $\bar{\nu}_e\to\bar{\nu}_e$  in JUNO experiment, as a function of neutrino energy $E$ in 
the scenarios of different CP phase combination, in which the brown line corresponds to the results in $(3+1)$ scheme while the blue one correspond to the standard $3\nu$ case.
}
\label{JUNO1}
\end{center}
\end{figure}

Around a nuclear power plant (NPP), there are plenty of antielectron neutrinos produced via $\beta$ decay
in nuclear reactions. Detectors can be put in suitable places near to the nuclear plant to explore reactor neutrino events.
Usually the baselines of such kind of experiments are in the range of short or medium baseline. 
For the exploring experiments, matter effect is not taken within the main considerations.
But the situation could be changed in precise measurement as experiment sensitivity may be affected.
The ongoing 
Jiangmen Underground Neitrino Observatory (JUNO) experiment \cite{An:2015jdp}, with its baseline $L=52.5\,\rm{km}$, is
one of such kind of experiments. 

Regarding to matter effect, whether the oscillation probability will change with  or without matter effect, both in purely 3 flavor active
neutrinos case and in the framework of active plus sterile neutrino case, is what we are concerned.
In Fig. \ref{JUNO1} we take the relative difference for probability, from matter to vacuum, and plot it varying by energy.
In order to  discriminate the Dirac phases' effect, we choose typic values of the two phases $\delta_{13}, \delta_{14}$ 
and make various combinations. In this analysis only
normal hierarchy (NH) of neutrino mass  situation is presented,
while the inverted hierarchy (IH) case has similar behaviors, though not shown explicitly. 
\begin{itemize}
\item[1)] Around the most promising range $E\sim 3\rm{MeV}$, the relative difference can reach  $4\%$,
which could possibly be distinguished by JUNO detector. 
\item [2)] The sterile neutrino contribution does not affect probability curve dramatically, that is to say
for short/medium baseline experiment, sterile neutrino effect is quite limited.
\item [3)] The effect from Dirac phase seems bleak, no distinction can be reflected from different phase combinations.
\end{itemize}
Hence we may conclude that 
the short and medium baseline experiments are not sensitive to the matter effect 
of sterile neutrino, as well as the CP-violating Dirac phases.

\subsection{Long baseline experiment}

Neutrino beam produced from accelerators usually carries higher energy and can be detected in a long distance from source. 
 In this part, we will take NO$\nu$A experiment \cite{Ayres:2007tu}, with its baseline $L=810\rm{km}$, 
 as an example to illustrate the properties of long baseline experiments case.

\begin{figure}[h]
\begin{tabular}{cc}
\includegraphics*[width=0.45\textwidth]{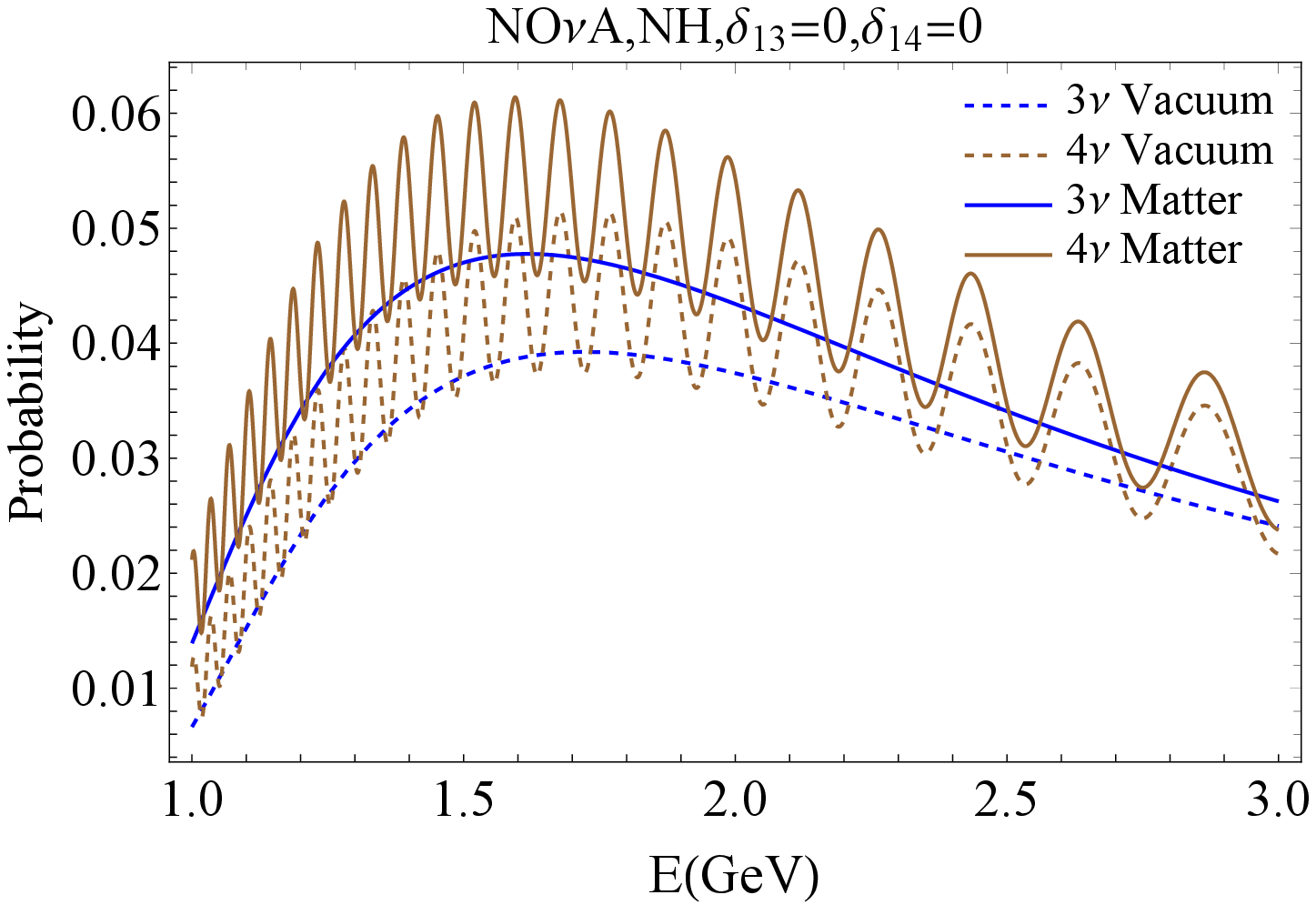}
&
\includegraphics*[width=0.45\textwidth]{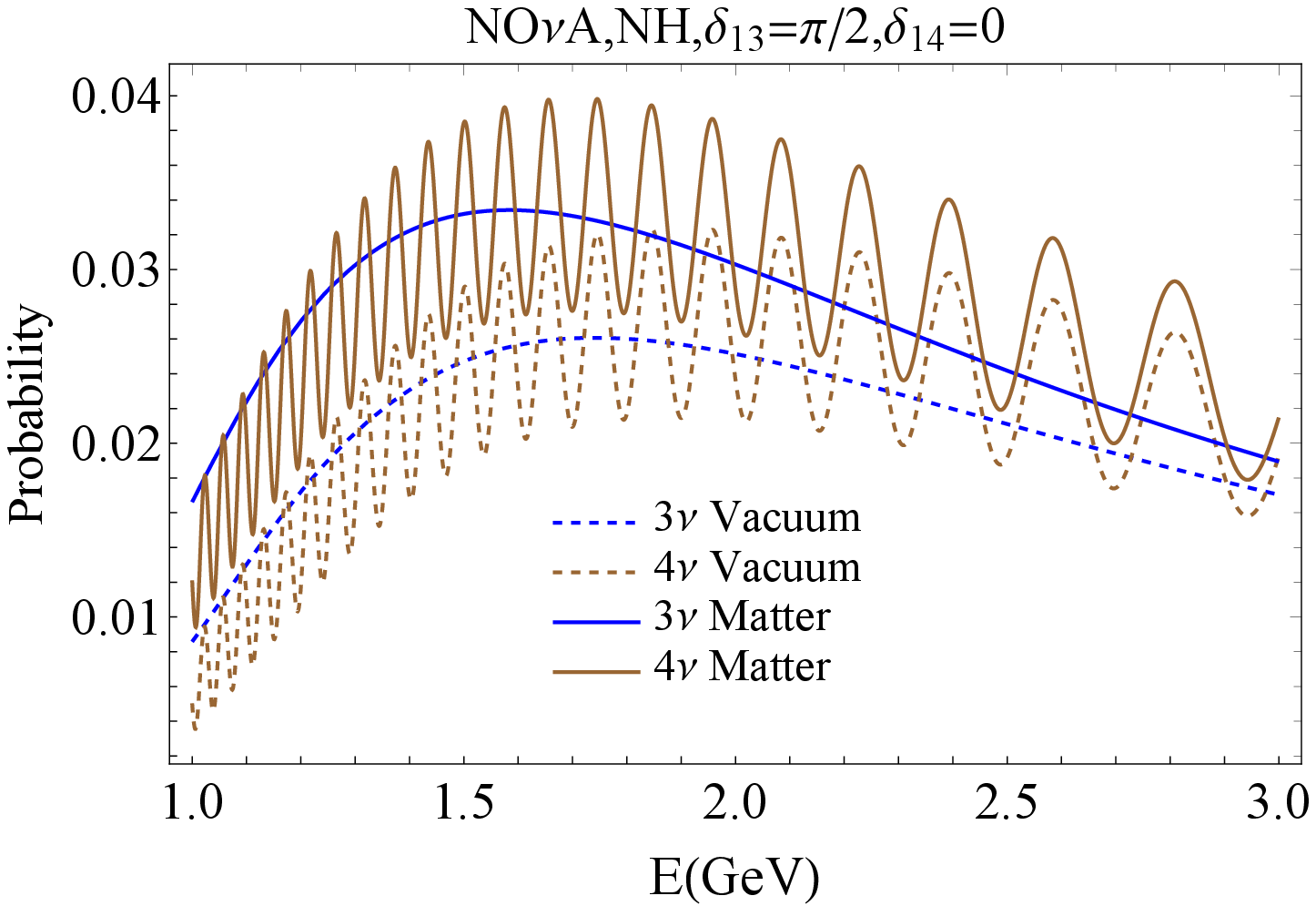}
\\
(a) & (b)\\
\\
\includegraphics*[width=0.45\textwidth]{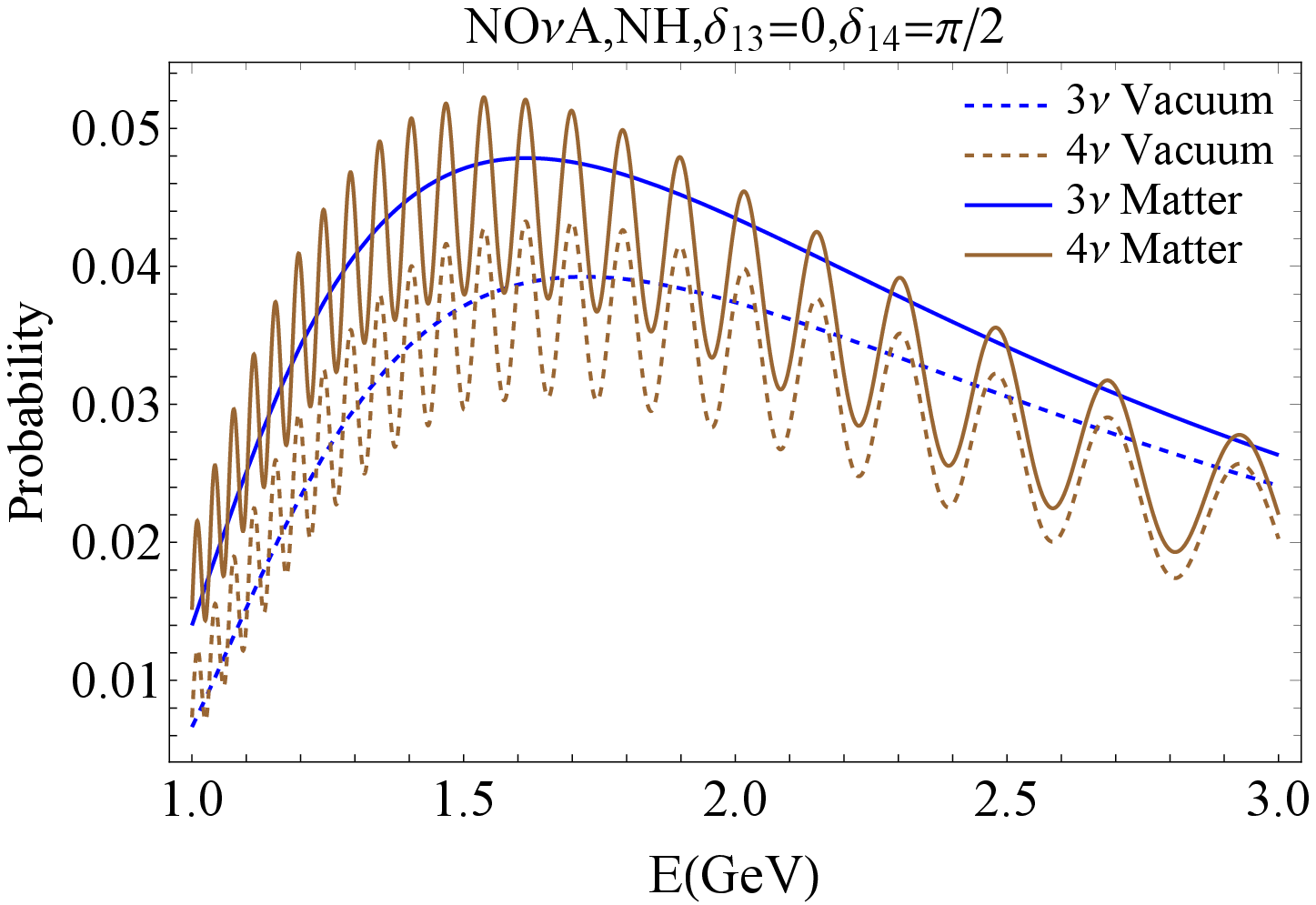}
&
\includegraphics*[width=0.45\textwidth]{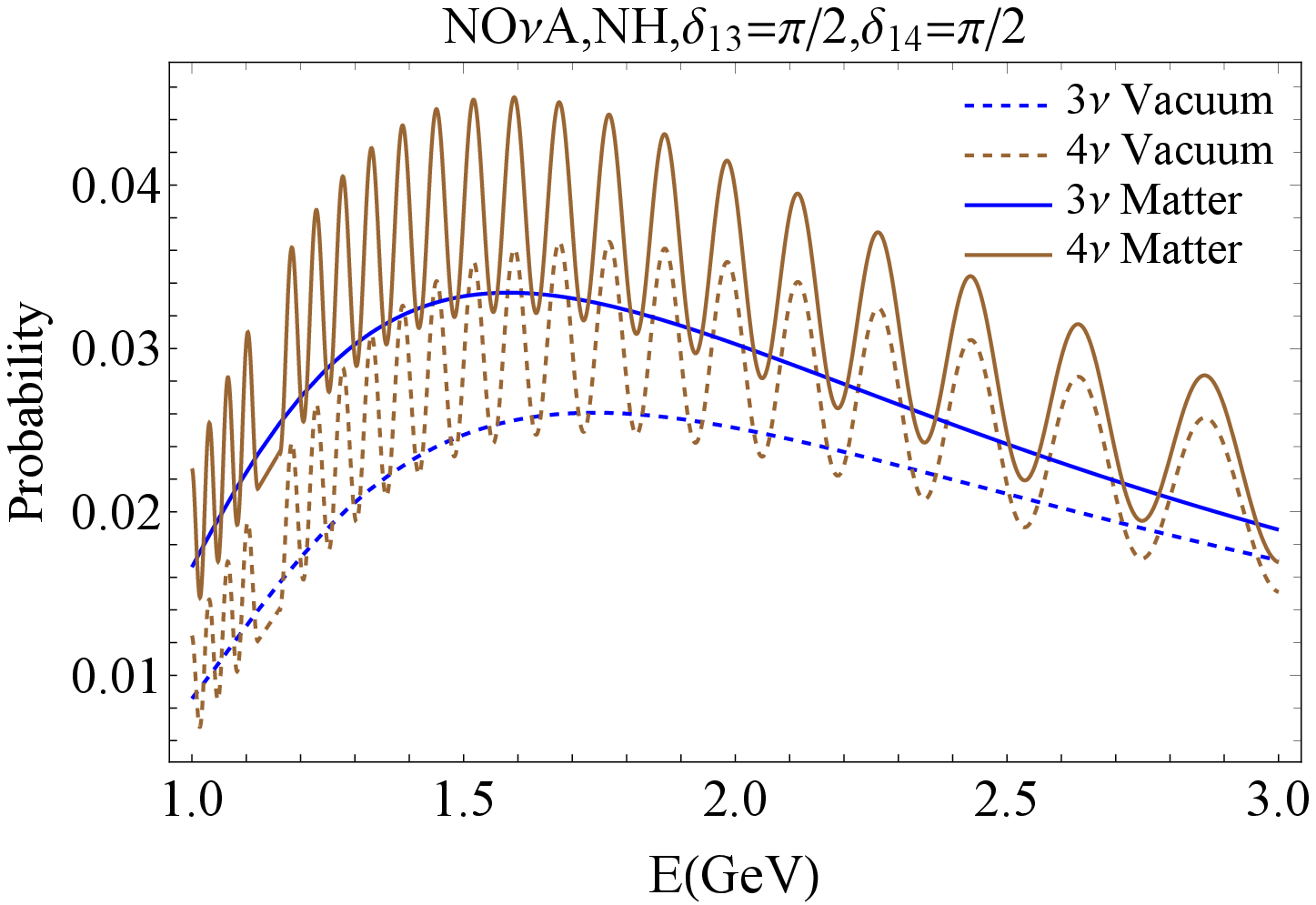}\\
(c) & (d) \\
\end{tabular}
\newline
\vspace{-0.6cm} \caption{Netutrino oscillation probabilities for the $\nu_\mu\to\nu_e$ channel as a function of energy $E$ for various dirac CP phase $\delta_{13},\delta_{14}$ in NO$\nu$A experiment. The blue (brown) curves denote the $3\nu(4\nu)$ cases, while the solid (dashed) curves correspond to the matter (vacuum) results.}
\label{Fig:NOvA1}
\end{figure}

Here we show the oscillation probability of appearance mode $\nu_\mu\to\nu_e$ in long baseline accelerator neutrino experiment 
in Fig. \ref{Fig:NOvA1}, where the brown curves stands for oscillation in 3+1 scheme and blue curves correspond to SM case while
solid (dashed) lines mean matter effect has (not) been contained.
In the plot, we have 
chosen typical CP phase combination of ($\delta_{13}, \delta_{14}$) in NH case, and consider its variation in energy range $1\sim 3 \rm{GeV}$. 
One can address the following points:
\begin{itemize}
\item[1)] The matter effect can not be negligible,  on the contrary, it is important both in $3\nu$ and $4\nu$ case.
At about $1~\rm{GeV}$ range, the relative difference for probabilities can be as large as $50\%$ in whichever scenario.
This difference could be $\sim 20\%$ around the maxima of oscillation probabilities.

\item[2)] No matter propagating in vacuum or in matter, sterile neutrino gives nonnegligible contribution to oscillation probability.
In each graph, the dashed lines have obvious deviation from their  solid correspondence.

\item[3)] The CP-violating Dirac phases also plays a non-ignorable role. 
By comparing Fig. 2a with Fig. 2c, one may see the oscillation probability has been 
affected.
In the scenario of $(\delta_{13}, \delta_{14})=(0,\frac{\pi}{2})$, one can see the blue lines are almost in the middle 
of corresponding brown lines.
But the blue curves has a distinct deviation from the average lines of brown ones 
in the scenario of  $(\delta_{13}, \delta_{14})=(\frac{\pi}{2},\frac{\pi}{2})$.

\end{itemize}


Therefore we may conclude that in the long baseline experiment, in the existence of sterile neutrino, the matter effect 
can not be ignored. 
The CP-violating Dirac phases in the mixing matrix may play an important role
in sterile neutrino's matter effect.
A more comprehensive analysis 
to display the entanglement of the phases is necessary, and  we will
show it in other places.


\section{Conclusion}
\label{sec:conclusion}
In this work, we have derived  exact formulas of oscillation probabilities with matter in medium and long baseline experiment  in presence of an additional light sterile neutrino. In particular, the key quantities contributing to oscillation probability, $\Delta\tilde{m}_{ij}^2$ and $\tilde{U}_{\alpha i}\tilde{U}_{\beta i}^*$, are shown explicitly. Based on exact formulas, we perform a detailed study of the matter effect correction in medium and long baseline experiments. We found that in medium baseline experiment, like $\rm{JUNO}$, the matter effect contribution is  negligible even in presence of light sterile neutrino. But in the long baseline experiment, taking
NO$\nu$A as an example,  the matter effect contribution plays a very important role especially when baseline grows.

\acknowledgments
J. Ling acknowledges the support by National Key R$\&$D program of China under  Grant No. 2018YFA0404103, 
NSFC (National Natural Science Foundation of China) under Grant No. 11775315 and young over-seas high-level talents introduction plan. 
F. Xu is supported partially by NSFC  under Grant No. 11605076, as well as the FRFCU (Fundamental Research Funds for the Central Universities in China) under the Grant No. 21616309.

\newpage

\appendix
\section{The parameterization of mixing matrix}
\label{sec:app-mix}
In $(3+1)$ scenario, the full neutrino mixing is characterized by a $4\times4$ matrix. To parameterize it, we need $6$ rotation angles and $3$ addtional Dirac phase angles.\cite{Fritzsch:1999ee} The Majorana phase angles are closed here because it doesn't involve in the oscillation process.
The mixing matrix can be constructed by $6$ two-dimensional rotations
\begin{equation}
U=R_{34}(\theta_{34},\delta_{34})\cdot R_{24}(\theta_{24})\cdot
R_{14}(\theta_{14},\delta_{14})\cdot
R_{23}(\theta_{23})\cdot R_{13}(\theta_{13},\delta_{13}) \cdot
R_{12}(\theta_{12})
\end{equation}
$R_{ij}$ is a four dimensional rotation matrix, and in the $(i,j)$ sublocks its elements reads
\begin{equation}
R_{ij}(\theta_{ij}, \delta)=\left(\begin{array}{cc}
c_{ij}& s_{ij} e^{-i\delta}\\
-s_{ij} e^{i\delta} & c_{ij}
\end{array}\right)
\end{equation}
with $c_{ij}=\cos\theta_{ij}, s_{ij}=\sin\theta_{ij}$.
The detailed matrix elements are
\begin{align}
&U_{e1}=c_{14} c_{13} c_{12}\\
&
U_{\mu1}= (-s_{24} s_{14} c_{13}  e^{i\delta_{14}}-c_{24}s_{23}s_{13} e^{i\delta_{13}})c_{12}
-c_{24}c_{23}s_{12}\nonumber\\
&U_{\tau 1}=c_{12}\left[ -s_{34} c_{24} s_{14}c_{13} e^{-i(\delta_{34}-\delta_{14})}
-s_{13} e^{i\delta_{13}}(c_{34}c_{23}-s_{34}s_{24}s_{23} e^{-i\delta_{34}})\right]\nonumber\\
&\hspace{1cm}
+s_{12}(s_{34}s_{24}c_{23} e^{-i\delta_{34}}+c_{34} s_{23})\nonumber\\
&U_{s1}=c_{12} \left[-c_{34}c_{24}s_{14} c_{13} e^{i\delta_{14}}
+s_{13} e^{i\delta_{13}}(s_{34}c_{23} e^{i\delta_{34}}+c_{34}s_{24}s_{23})\right]\nonumber\\
&\hspace{1cm}
-s_{12}(-c_{34}s_{24}c_{23}+s_{34}s_{23} e^{i\delta_{34}})\nonumber\\
&U_{e2}=c_{14}c_{13}s_{12}\nonumber\\
&U_{\mu 2}=c_{24}c_{23}c_{12}-s_{12}(s_{24}s_{14}c_{13}e^{i\delta_{14}}
+c_{24}s_{23}s_{13}e^{i\delta_{13}})\nonumber\\
&U_{\tau 2}=-c_{12}(s_{34} s_{24} c_{23} e^{-i\delta_{34}}+c_{34}s_{23})\nonumber\\
&\hspace{1cm}
-s_{12}\left[s_{34}c_{24}s_{14}s_{13} e^{-i(\delta_{34}-\delta_{14})}
+s_{13} e^{i\delta_{13}} (c_{34} c_{23}-s_{34}s_{24}s_{23} e^{-i\delta_{34}})\right]\nonumber\\
& U_{s2}=c_{12}\left(-c_{34}s_{24}c_{23}+s_{34}s_{23} e^{i\delta_{34}}\right)\nonumber\\
&\hspace{1cm}
-s_{12}\left[ c_{34}c_{24}s_{14}c_{13}e^{i\delta_{14}}-s_{13}e^{i\delta_{13}}\left(
s_{34}c_{23}e^{i\delta_{34}}+c_{34}s_{24}s_{23}\right)\right]\nonumber\\
&U_{e3}=c_{14}s_{13} e^{-i\delta_{13}}\nonumber\\
&U_{\mu 3}=c_{24}s_{23}c_{13}-s_{24}s_{14}c_{13}e^{i(\delta_{14}-\delta_{13})}\nonumber\\
&U_{\tau 3}=c_{13}\left(c_{34}c_{23}-s_{34}s_{24}s_{23}e^{-i\delta_{34}}\right)-s_{34}c_{24}s_{14}s_{13}e^{-i(\delta_{34}
-\delta_{14}+\delta_{13})}\nonumber\\
&U_{s3}=-c_{13}\left(s_{34}c_{23}e^{i\delta_{34}}
+c_{34}s_{24}s_{23}\right)
-c_{34}c_{24}s_{14}s_{13}e^{i(\delta_{14}-\delta_{13})}\nonumber\\
&U_{e4}=s_{14} e^{-i\delta_{14}}\nonumber\\
&U_{\mu 4}=s_{24}c_{14}\nonumber\\
&U_{\tau 4}=s_{34}c_{24}c_{14}e^{-i\delta_{34}}\nonumber\\
&U_{s4}=c_{34}c_{24}c_{14}.\nonumber
\end{align}
Apparently if we close the angles related the forth neutrino, this $4\times 4$ lepton mixing matrix will
reduce to $3\times3$ PMNS matrix.

\section{Calculation of effective mass-square difference $\hat\Delta m_{i1}^2$}
\label{subsec:eigenequation}
In order to solve Eq.(\ref{eq:diagH}), we resort to solving a quartic equation in below,
where its root is denoted as $\lambda_i(i=1,2,3,4)$.
\begin{equation}
\lambda^4+b\lambda^3+c\lambda^2+d\lambda+e=0,
\end{equation}
and the coefficients are defined as
\begin{align}\label{coefficent1}
b&=-\sum_i\Delta m_{i1}^2-A-A'\notag\\
c&=A\sum_i\Delta m_{i1}^2\Big(1-|U_{ei}|^2\Big)+A'\sum_i\Delta m_{i1}^2\Big(1-|U_{si}|^2\Big)+A{A'}+\sum_{i<j}\Delta m_{i1}^2\Delta m_{j1}^2\notag\\
d&=A{A'}\sum_i\Delta m_{i1}^2\Big(|U_{ei}|^2+|U_{si}|^2-1\Big)-\sum_{i,j,k}\frac{\varepsilon_{ijk}^2}{2}\Delta m_{i1}^2\Delta m_{j1}^2\Big(A|U_{ek}|^2+{A'}|U_{sk}|^2\Big)-\Delta m_{21}^2\Delta m_{31}^2\Delta m_{41}^2\notag\\
e&=A{A'}\sum_{i,j,k,l}\frac{\varepsilon_{ijkl}^2}{4}\Delta m_{i1}^2\Delta m_{j1}^2|U_{ek}U_{sl}-U_{el}U_{sk}|^2+\Delta m_{21}^2\Delta m_{31}^2\Delta m_{41}^2\Big({A'}|U_{s1}|^2+A|U_{e1}|
^2\Big)
\end{align}
with Levi-Civita symbol $\varepsilon_{ijk}$ and $\varepsilon_{ijkl}$. More  auxiliary qualities are introduced to make the result more concise
\begin{align}\label{coefficent2}
p&=\frac{8c-3b^2}{8},\qquad\qquad q=\frac{b^3-4bc+8d}{8}\notag\\
K_0&=c^2-3bd+12e,\qquad K_1=2c^3-9bcd+27b^2e+27d^2-72ce\notag\\
S&=\frac{1}{2}\sqrt{-\frac{2}{3}p+\frac{2}{3}\sqrt{K_0}\cos\frac{\phi}{3}},\qquad\phi=\arccos\left(\frac{K_1}{2K_0^{3/2}}\right).
\end{align}
With the above notations,  we find solutions  of $\lambda_i$
\begin{align}\label{lambda}
\lambda_1&=-\frac{b}{4}-S-\frac{1}{2}\sqrt{-4S^2-2p+\frac{q}{S}}\notag\\
\lambda_2&=-\frac{b}{4}-S+\frac{1}{2}\sqrt{-4S^2-2p+\frac{q}{S}}\notag\\
\lambda_3&=-\frac{b}{4}+S-\frac{1}{2}\sqrt{-4S^2-2p-\frac{q}{S}}\notag\\
\lambda_4&=-\frac{b}{4}+S+\frac{1}{2}\sqrt{-4S^2-2p-\frac{q}{S}}
\end{align}
where $\lambda_1<\lambda_2<\lambda_3<\lambda_4$. Notice that we don't assume the a normal or inverted mass hierachy. Thus eq.(\ref{lambda}) and the inequality above are hierachy independent. If neutrinos are in the inverted mass hierachy $(m_3<m_1<m_2<m_4)$, we can simply set $\tilde{m}_3^2=m_1^2+\lambda_1,\tilde{m}_1^2=m_1^2+\lambda_2,\tilde{m}_2^2=m_1^2+\lambda_3,\tilde{m}_4^2=m_1^2+\lambda_4$. If they are in the normal mass hierachy $(m_1<m_2<m_3<m_4)$, the result is the eq.(\ref{massdiff})

\section{Calculation of effective matrix $\tilde{U}_{\alpha i}\tilde{U}^*_{\beta i}$}
\label{app:mixing}

From Eq. (\ref{eq:P-mass}), we may see the oscillation probability depends on some certain combinations of entries of $\tilde{U}$, that is 
$\tilde{U}_{\alpha i}\tilde{U}^*_{\beta i}$, where $i$ is not summed. 
In \cite{Zhang:2006yq} some of the calculations has been done. The calculation of $\alpha=\beta$ case can be confirmed, however, the
result for $\alpha\neq \beta$ seems wrong.  Thus a reconsideration is required. 
In this section, we will complete this mission by an explicit calculation.

Since both $\tilde{U}$ and $U$ are unitary, one can subtract a diagonal matrix from Eq. (\ref{effectiveH}) and Eq. (\ref{effectiveH}), 
leading to
{\footnotesize{
\begin{equation}
\tilde{U}\left(\begin{array}{cccc}
0 & & & \\
 & \Delta\tilde{m}^2_{21} & &\\
 & & \Delta\tilde{m}^2_{31} & \\
 & & & \Delta\tilde{m}^2_{41}
 \end{array}\right)\tilde{U}^\dagger=	
 {U}\left(\begin{array}{cccc}
0 & & & \\
 & \Delta{m}^2_{21} & &\\
 & & \Delta{m}^2_{31} & \\
 & & & \Delta{m}^2_{41}
 \end{array}\right){U}^\dagger
 +
 \left(\begin{array}{cccc}
A-\hat{\Delta}m_{11}^2& & & \\
 &-\hat{\Delta}m_{11}^2 & &\\
 & & -\hat{\Delta}m_{11}^2 & \\
 & & & A'-\hat{\Delta}m_{11}^2
 \end{array}\right).\label{one}
\end{equation}
}}
We can further write down 
\begin{equation}
\sum_{i=2}^4\Delta\tilde{m}_{i1}^2\tilde{U}_{\alpha i}\tilde{U}_{\beta i}^*=\sum_{i=2}^4\Delta m_{i1}^2U_{\alpha i}U_{\beta i}^*+\mathcal{A}_{\alpha\beta}
-\hat{\Delta}m^2_{11} \delta_{\alpha\beta}
\label{two}
\end{equation}
with {$\mathcal{A}_{\alpha\beta}\equiv A\delta_{e\alpha}\delta_{e\beta}+A'\delta_{s\alpha}\delta_{s\beta}$.}
Similarly by taking the square and cube of Eq. (\ref{one}), we can have equations corresponding to  Eq. (\ref{two})
\begin{align}\label{relation1}
\sum_{i=2}^4(\Delta\tilde{m}_{i1}^2)^2\tilde{U}_{\alpha i}\tilde{U}_{\beta i}^*&=\Big(\sum_{i=2}^4\Delta m_{i1}^2U_{\alpha i}U_{\beta i}^*+\mathcal{A}_{\alpha\beta}
-\hat{\Delta}m^2_{11} \delta_{\alpha\beta}
\Big)^2\notag\\
\sum_{i=2}^4(\Delta\tilde{m}_{i1}^2)^4\tilde{U}_{\alpha i}\tilde{U}_{\beta i}^*&=\Big(\sum_{i=2}^4\Delta m_{i1}^2U_{\alpha i}U_{\beta i}^*+\mathcal{A}_{\alpha\beta}
-\hat{\Delta}m^2_{11} \delta_{\alpha\beta}
\Big)^3
\end{align}
{The square and cube on the right-handed side of (\ref{relation1}) means $M^2=M_{\alpha\beta}^2=\sum\limits_\gamma M_{\alpha\gamma}M_{\gamma\beta},M^3=M_{\alpha\beta}^3=\sum\limits_{\gamma,\rho}M_{\alpha\gamma}M_{\gamma\rho}M_{\rho\beta}$}
The above equations can be written formally in a much more compact matrix, as
\begin{equation}
\begin{pmatrix}
\Delta\tilde{m}_{21}^2&\Delta\tilde{m}_{31}^2&\Delta\tilde{m}_{41}^2\\
(\Delta\tilde{m}_{21}^2)^2&(\Delta\tilde{m}_{31}^2)^2&(\Delta\tilde{m}_{41}^2)^2\\
(\Delta\tilde{m}_{21}^2)^4&(\Delta\tilde{m}_{31}^2)^4&(\Delta\tilde{m}_{41}^2)^4
\end{pmatrix}
\begin{pmatrix}
\tilde{U}_{\alpha2}\tilde{U}_{\beta2}^*\\
\tilde{U}_{\alpha3}\tilde{U}_{\beta3}^*\\
\tilde{U}_{\alpha4}\tilde{U}_{\beta4}^*
\end{pmatrix}
=
\begin{pmatrix}
\sum\limits_{i=2}^4\Delta m_{1i}^2U_{\alpha i}U_{\beta i}^*+\mathcal{A}_{\alpha\beta}-\hat{\Delta}m^2_{11} \delta_{\alpha\beta}\\
\Big(\sum\limits_{i=2}^4\Delta m_{1i}^2U_{\alpha i}U_{\beta i}^*+\mathcal{A}_{\alpha\beta}\Big)^2-\hat{\Delta}m^2_{11} \delta_{\alpha\beta}\\
\Big(\sum\limits_{i=2}^4\Delta m_{1i}^2U_{\alpha i}U_{\beta i}^*+\mathcal{A}_{\alpha\beta}\Big)^3-\hat{\Delta}m^2_{11} \delta_{\alpha\beta}
\end{pmatrix}.
\label{keyequation}
\end{equation}
By solving linear equation Eq.(\ref{keyequation})
straightforwardly,  we may get two classes of solutions shown below,
denoted as $\alpha=\beta$ and $\alpha\neq \beta$.
\begin{itemize}
\item[1)] $\alpha=\beta$

The solution is
\begin{equation}\label{Ueq}
|\tilde{U}_{\alpha i}|^2=\frac{1}{\prod\limits_{k\neq i}\Delta\tilde{m}_{ik}^2}\sum_{j=1}^4\left(F_\alpha^{ij}+C_\alpha\right)|U_{\alpha j}|^2
\end{equation}
with the auxiliary functions
\begin{equation}
F_\alpha^{ij}=\prod\limits_{k\neq i}\left(\mathcal{A}_{\alpha\alpha}+\Delta m_{j1}^2-\hat{\Delta}m_{k1}^2\right),\qquad C_\alpha=-\frac{1}{2}\sum_{\gamma,m,n}(\Delta m_{mn}^2)^2U_{\alpha m}U_{\gamma n}U_{\gamma m}^*U_{\alpha n}^*\mathcal{A}_{\gamma\gamma}
\end{equation}

\item[2)] $\alpha\neq \beta$

Solution in this case is obtained as
\begin{align}
\tilde{U}_{\alpha i}\tilde{U}_{\beta i}^*&=\frac{1}{\prod\limits_{k\neq i}\tilde{\Delta}m_{ik}^2}\Bigg(\sum_{j=2}^4F_{\alpha\beta}^{ij}U_{\alpha j}U_{\beta j}^*+C_{\alpha\beta}\Bigg)\,\hspace{2cm} (i=2,3,4)\label{U234}\\
\tilde{U}_{\alpha1}\tilde{U}_{\beta1}^*&=
-\tilde{U}_{\alpha2}\tilde{U}_{\beta2}^*-\tilde{U}_{\alpha3}\tilde{U}_{\beta3}^*-\tilde{U}_{\alpha4}\tilde{U}_{\beta4}^*\label{U1},
\end{align}
in which we have especially introduced two important quantities,
\begin{align}
F_{\alpha\beta}^{ij}=&\Bigg[(\mathcal{A}_{\alpha\alpha}^2+\mathcal{A}_{\beta\beta}^2+\mathcal{A}_{\alpha\alpha}\mathcal{A}_{\beta\beta})\Delta m_{j1}^2+(\mathcal{A}_{\alpha\alpha}+\mathcal{A}_{\beta\beta})\Delta m_{j1}^2\Big(\Delta m_{j1}^2-\sum_{k\neq i}\hat{\Delta}m_{k1}^2\Big)\notag\\
&+(\Delta m_{j1}^2)^3{
-\sum_{k\neq i}(\Delta m_{j1}^2)^2\hat{\Delta}m_{k1}^2
}
+\sum_{k,l;k\neq l\neq i}\Delta m_{j1}^2\hat{\Delta}m_{k1}^2\hat{\Delta}m_{l1}^2\Bigg]
\label{eq:F}\\
C_{\alpha\beta}=&A'\sum_{k,l}\Delta m_{k1}^2\Delta m_{l1}^2U_{\alpha k}U_{\beta l}^*U_{sk}U_{sl}^*+A\sum_{k,l}\Delta m_{k1}^2\Delta m_{l1}^2U_{ek}^*U_{el}U_{\alpha k}U_{\beta l}^*\label{eq:C}
\end{align}
In particular the second equation Eq. (\ref{U1}) is obtained by making use of the unitary property of $\tilde{U}$. 
\end{itemize}

As a summary, we may claim that Eq. (\ref{Ueq} ) together with Eq. (\ref{U234}) and Eq. (\ref{U1})  are the most general expressions for $\tilde{U}_{\alpha i}\tilde{U}^*_{\beta i}$.


\end{document}